\documentclass[sigconf,screen,nonacm]{acmart}

\AtBeginDocument{%
  \providecommand\BibTeX{{%
    \normalfont B\kern-0.5em{\scshape i\kern-0.25em b}\kern-0.8em\TeX}}}

\usepackage{textgreek}
\usepackage{longtable}
\usepackage{amsmath,amssymb,amsfonts}
\usepackage{algorithm}
\usepackage{textcomp}
\usepackage{xcolor}
\usepackage{soul}
\usepackage{booktabs}
\usepackage{microtype}
\usepackage{rotating}
\usepackage{graphicx}
\usepackage{paralist}
\usepackage{tabularx}
\usepackage{multicol}
\usepackage{multirow}
\usepackage{pbox}
\usepackage{enumitem}	
\usepackage{colortbl}
\usepackage{xspace}
\usepackage{url}
\usepackage{tabularx}
\usepackage{lscape}
\usepackage{listings}
\usepackage{color}
\usepackage{comment}
\usepackage{soul}
\usepackage{multibib}
\usepackage{array}
\usepackage{tcolorbox}

\newcommand{\ie}{\textit{i.e.,}\xspace}
\newcommand{\eg}{\textit{e.g.,}\xspace}

\newcommand{\etc}{\textit{etc.}\xspace}
\newcommand{\etal}{\textit{et al.}\xspace}

\newcommand{\secref}[1]{Section~\ref{#1}\xspace}
\newcommand{\figref}[1]{Figure~\ref{#1}\xspace}
\newcommand{\tabref}[1]{Table~\ref{#1}\xspace}
\newcommand{\tool}[1]{{\sc #1}\xspace}

\newcommand{\RQ}[1]{$RQ_{#1}$\xspace}

\newcommand\new[1]{\textcolor{black}{#1}}

\begin{document}
\sloppy 
\title{On the Relationship between Refactoring Actions and Bugs:\\A Differentiated Replication}

\author{Massimiliano Di Penta}
\email{dipenta@unisannio.it}
\affiliation{%
  \institution{University of Sannio}
  \streetaddress{Via Traiano, 9}
  \city{Benevento}
  \country{Italy}
  \postcode{82100}
}

\author{Gabriele Bavota}
\email{gabriele.bavota@usi.ch}
\affiliation{%
  \institution{Universit\`a della Svizzera italiana}
  \streetaddress{P.O. Box 1212}
  \city{Lugano}
  \country{Switzerland}
  \postcode{XXX}
}

\author{Fiorella Zampetti}
\email{fiorella.zampetti@unisannio.it}
\affiliation{%
   \institution{University of Sannio}
  \streetaddress{Via Traiano, 9}
  \city{Benevento}
  \country{Italy}
  \postcode{82100}
}

\renewcommand{\shortauthors}{Di Penta et al.}

\begin{abstract}
Software refactoring aims at improving code quality while preserving the system's external behavior. Although in principle refactoring is a behavior-preserving activity, a study presented by Bavota \etal in 2012 reported the proneness of some refactoring actions (\eg pull up method) to induce faults. The study was performed by mining refactoring activities and bugs from three systems. Taking profit of the advances made in the mining software repositories field (\eg better tools to detect refactoring actions at commit-level granularity), we present a differentiated replication of the work by Bavota \etal in which we (i) overcome some of the weaknesses that affect their experimental design, (ii) answer the same research questions of the original study on a much larger dataset (3 \emph{vs} 103 systems), and (iii) complement the quantitative analysis of the relationship between refactoring and bugs with a qualitative, manual inspection of commits aimed at verifying the extent to which refactoring actions trigger bug-fixing activities. The results of our quantitative analysis confirm the findings of the replicated study, while the qualitative analysis partially demystifies the role played by refactoring actions in the bug introduction.
\end{abstract}




\definecolor{gray}{rgb}{0.83,0.83,0.83}
\definecolor{green}{rgb}{0.56,0.93,0.56}

\newtcolorbox{resultbox}{colback=gray, arc=0.5mm, top=0.5mm, bottom=0.5mm, left=0.5mm, right=0.5mm}

\maketitle
\eject
%%%%%%%%%%%%%%%%%%%%%%%%%%%%%%%%
%%%%%%%%%%%%%%%%%%%%%%%%%%%%%%%%
\section{Introduction}
\label{sec:intro}
%%%%%%%%%%%%%%%%%%%%%%%%%%%%%%%%
%%%%%%%%%%%%%%%%%%%%%%%%%%%%%%%%

Software refactoring has been extensively studied by the research community, through empirical studies investigating how and why developers perform refactoring \cite{Wang:icsm2009,Emerson:tse2011,Kim:fse2012,Silva:fse2016,Peruma:iwor2018,Vassallo:scp2019}, how refactoring relates with other development tasks  (\eg merge conflicts \cite{Mahmoudi:saner2019}), with software quality indicators (\eg quality metrics) \cite{Stroggylos:2007,szoke2014bulk,Alshayeb20091319,Chavez:sbes2017}, and with developers' productivity \cite{Moser:2008}. Some studies (\eg Kim \etal \cite{Kim:fse2012}) indicated that often developers are concerned about performing refactoring activities as it may cause the introduction of bugs. 

The relationship between refactoring and bugs has been the subject of several studies, that analyzed software repositories to understand the extent to which refactoring activities introduce bugs \cite{Weissgerber:msr2006,BavotaCLPOS12,Ferreira:icse2018}. Wei{\ss}gerber and Diehl \cite{Weissgerber:msr2006} studied the correlation between refactoring activities and bug reports opened in the subsequent days, finding no strong correlation. However, their study did not link refactoring activities in a specific file with bug-fixes performed on that same file. 

\new{In a previous work, some of the authors\footnote{In the following we refer previous work as Bavota \etal because the set of authors only partially overlaps.}
 \cite{BavotaCLPOS12}} presented a study overcoming this limitation, showing that refactoring actions involving hierarchies (\eg {\em push-down method}) induce bug-fixing commits more frequently than other refactoring types. They used the \tool{Ref-Finder} \cite{PreteRSK10} tool to create a dataset of 12,922 manually-validated refactoring actions, detected comparing subsequent releases (63 in total) of three Java systems. By comparing releases, Bavota \etal \cite{BavotaCLPOS12} assumed that a specific refactoring was performed on a file $F_j$ between  releases $R_i$ and $R_{i+1}$ of a given system, while the exact refactoring-related commit was unknown. Then, by mining the change history of the three systems, the authors identified bug-fixing commits by linking commit messages and issue tracker data using a keyword-based approach \cite{FischerPG03} (\eg ``\emph{fixed issue \#ID}'', where ID was the id of an issue on the issue tracker of the mined system).
 Finally, for each bug-fixing commit, they identified its fix-inducing commits using the SZZ algorithm \cite{Sliwerski:msr2005}. 
 Using such data, Bavota \etal assumed that a refactoring action performed on file $F_j$ between $R_i$ and $R_{i+1}$ induced a fix if a bug-inducing commit $c$ identified by the SZZ was performed on $F_j$ between $R_i$ and $R_{i+1}$. Thus, there is a strong assumption made in the experimental design: Since the refactoring actions were captured between releases, it is not possible to know whether the refactoring was actually implemented in the bug-inducing commit $c$. Also, some refactoring actions may not be detected because of the large differences that may occur between two releases. 
 
This, together with the small size (three projects) are the main limitations of this study. 

More recently, Ferreira \etal \cite{Ferreira:icse2018} reported preliminary results of a mining-based study performed on five systems and overcoming the main design issue of the work by Bavota \etal~\cite{BavotaCLPOS12}. Ferreira \etal mined both refactoring actions and bug-inducing changes at commit-level, looking for how ``close'' the refactoring actions were to bug-inducing changes. They confirmed the relationship between refactoring actions and bugs showing, however, that many bugs are not the direct consequence of the refactoring action, but of changes implemented later on the refactored code. By using a tool-chain similar to the one adopted by Ferreira \etal \cite{Ferreira:icse2018}, we present a differentiated replication of the study by Bavota \etal \cite{BavotaCLPOS12}. We overcome several limitations of that study by:

\emph{Taking profit of the recent advances made in the mining software repositories field}. This reflects in (i) better refactoring miner tools able to precisely identify refactoring actions at commit-level granularity \cite{Tsantalis:icse2018}, thus avoiding the  assumption made in the original study done at release-level; (ii) enhanced implementations of the SZZ algorithm, overcoming some of the limitations of the original algorithm \cite{CostaMSKCH17}; (iii) a line-level linking between refactoring actions and bug-fixing activities (as compared to the file-level linking done in previous studies).

\emph{Considering the possible impact of the size confounding factor on the achieved results}. While the original study indicated a relationship between specific refactoring actions and the introduction of bugs, the authors ignored the possible impact of the size confounding factor on this finding (\eg refactoring is usually performed in larger commits and larger commits are more likely to introduce bugs).

\emph{Complementing the quantitative analysis with a systematic qualitative evaluation}. We manually analyze a statistically significant sample of 384 commits identified as fix-inducing refactoring actions (\ie those that induced a bug-fixing activity) to study whether the performed refactoring actions actually induced the bug-fix. This analysis provides more confidence in the reported quantitative findings.

\emph{Answering the same research questions presented in \cite{BavotaCLPOS12}, but on a larger scale}. We answer the research questions presented in \cite{BavotaCLPOS12} both on the same three systems used in the original study, as well as, on a set of 100 open source Java projects. This increases the generalizability of the findings. \vspace{0.05cm}

Despite the different experimental design adopted, our quantitative analysis confirms most of the findings of the original study. However, we also unveil the significant role played by the size confounding factor in inducing bug-fixing activities. Also, our qualitative analysis shows that, while the SZZ can identify the commit implementing the refactoring(s) as the last one modifying the code then subject to bug-fixing activities, in many cases the bug was already in the system before the refactoring even happened. 

\new{The obtained results trigger further research in the area of automated refactoring, but also warns developers about possible risks associated with refactoring activities, if the latter are not accompanied by suitable verification \& validation.}

\new{\textbf{Paper structure.} \secref{sec:design} describes the study design. Results are discussed in \secref{sec:results}, while their threats to validity in \secref{sec:threats}. After a discussion of related work (\secref{sec:related}), \secref{sec:conclusions} concludes the paper.}
%\input{sections/original}
%%%%%%%%%%%%%%%%%%%%%%%%%%%%%%%%
%%%%%%%%%%%%%%%%%%%%%%%%%%%%%%%%
\section{Study Design}
\label{sec:design}
%%%%%%%%%%%%%%%%%%%%%%%%%%%%%%%%
%%%%%%%%%%%%%%%%%%%%%%%%%%%%%%%%

The \emph{goal} of the study is to perform a differentiated replication of the work by Bavota \etal \cite{BavotaCLPOS12}, in which the authors investigated the extent to which refactoring actions trigger bug-fixing activities. The \emph{context} is represented by the history of 103 Java projects, and in particular by the refactoring operations and bug-fixes performed by their developers. %The \emph{perspective} is that of researchers interested in (i) enlarge the empirical knowledge linking refactoring actions and bugs, and (ii) corroborate/contradict previous findings reported in the literature \cite{BavotaCLPOS12}.

\newcommand{\rqone}{Are refactoring-related commits more likely to induce fixes than other commits?}

\newcommand{\rqtwo}{To what extent is the relationship between refactoring actions and fix induced changes influenced by the effect of size?}

\newcommand{\rqthree}{What kinds of refactoring types are more likely to induce fixes?}

\newcommand{\rqfour}{To what extent does refactoring actually trigger bug-fixing activities?}

We address the following research questions (RQs):

\textbf{RQ$_1$} \emph{\textbf{\rqone}} This RQ mirrors the RQ$_1$ from the original work of Bavota \etal \cite{BavotaCLPOS12}. They answered this RQ by mining refactoring actions and fix-inducing changes performed between subsequent releases of three systems. %This means assuming that: (i) a certain refactoring has been performed between two releases $R_i$ and $R_{i+1}$ on file $F_j$ (but not in which commit this refactoring was performed); and (ii) that $F_j$ has been involved in fix-inducing commits (\ie commits that, accordingly to the SZZ algorithm induced a bug fix) between $R_i$ and $R_{i+1}$, again, without knowing whether one of these commits was also the one implementing the refactoring, but also with the possibility that, when comparing releases, some refactoring actions are not detected because of the large differences between releases. 
Using this data, Bavota \etal investigated whether refactoring operations are likely to induce bug-fixes. However, as also acknowledged \cite{BavotaCLPOS12}, the strong (unverified) assumption behind the study is that there is an overlap between the fix-inducing commits and the commits that implemented the refactoring actions. Instead of performing our replication at release-level, we use a commit-level granularity. This means that we know the exact commits in which refactoring operations have been performed in a specific file $F_j$ and, as a consequence, we can check whether those commits induced a fix or not. We also improved other aspects on top of the original experimental design. Finally, while we answer RQ$_1$ by using the same three systems adopted in the original study~\cite{BavotaCLPOS12}, we also answer RQ$_1$ in a large-scale study involving 100 open source projects.

\textbf{RQ$_2$} \emph{\textbf{\rqtwo}} Bavota \etal  did not consider the size of the code change as a possible confounding factor in their analysis. However, it is well-known that large commits (\ie commits impacting a large number of files/lines/code churns) have a higher probability of inducing a bug~\cite{KimWZ08}. It is possible that commits implementing refactoring operations are more likely to induce bug-fixes simply because they are larger than commits implementing other types of changes (\eg bug-fixes, enhancements). RQ$_2$ aims at investigating the role played by the commit size co-factor in the relationship between refactoring actions and fix-inducing changes.

\textbf{RQ$_3$} \emph{\textbf{\rqthree}} RQ$_3$ mirrors the RQ$_2$ of the original study, and analyzes the likelihood that different types of refactoring (\eg \emph{extract class}, \emph{pull up method}) trigger bug-fixing activities. 

\textbf{RQ$_4$} \emph{\textbf{\rqfour}} RQ$_4$ is a qualitative analysis we perform on a sample of the fix-inducing commits we identified in our quantitative study as responsible for both (i) implementing a refactoring, and (ii) inducing a bug-fixing activity. In other words, these should be the commits where there is a cause-effect relationship between refactoring and bug introduction. %We verify whether this is the case through a manual analysis, which was not performed in the original study~\cite{BavotaCLPOS12}.

\subsection{Context Selection}
\label{sec:context}
We answer our research questions by mining the change history of 103 projects. Three of them, namely Apache Ant, ArgoUML, and Apache Xerces-J, are the Java projects used in the replicated study~\cite{BavotaCLPOS12}, while the remaining 100 were selected from GitHub through the following procedure.

Our initial idea was to mine popular and large projects from GitHub, excluding forked projects, coding tutorials, and personal projects, as well as projects having less than 100 issues and 1,000 commits, to ensure the availability of a long change history to study. 

Also, we decided to ignore projects having less than 80\% of their code written in Java since the refactoring detector used in our study~\cite{Tsantalis:icse2018} only works with Java. Finally, since in our study it is of crucial importance to identify bug-fixing commits, we also wanted to exclude repositories not using a clear label for bugs and those not consistently referencing in commit notes the id(s) of the issue(s) closed by the commit. Concerning the first point (\ie label for bugs), in GitHub every project can define its own set of labels to ``tag'' the opened issues, thus indicating bugs, feature requests, \etc As for the second point, having an explicit link between commits and bugs allows to precisely identify the bug-fixing commits needed for our study. %Indeed, by applying the SZZ on the bug-fixing commits (details follow), we can identify fix-inducing commits, and verify the extent to which those match with commits implementing refactoring activities.

To this aim, we used the GitHub API \cite{githubAPI} to extract the list of projects having at least 100 issues and Java as their ``first language''. The latter criterion means that Java is the most used language in the project, but does not guarantee that the vast majority of the code is written in Java. %For example, a project may use five languages, with the most used being Java with 30\% of the overall code. 
Since the GitHub API returns at most 1,000 results per search, we generated several requests, each having a specific size range. We used the \texttt{size:min..max} argument to retrieve only projects within a specific size range. In this way, we increased the number of returned results to up 1,000 $\times$ $n$, where $n$ is the number of considered size ranges. %We searched for projects having a size ranging from 1,000 to 500,000 kilobytes, with a step of 1,000 kilobytes.
Note that, while such a search heuristic does not allow to identify all possible GitHub projects having at least 100 issues and Java as their primary language, this is not important for the sake of our study. Here the goal was to just collect a set of candidate projects that then we can manually validate to decide which ones to include in our study. We collected 2,538 projects, and two of the authors inspected them to check the selection criteria previously mentioned. %clear label to identify bugs, consistent use of issue references in commit notes, at least 80\% of Java code, at least 1,000 commits. 
After analyzing the first 1,000 projects (by sorting them in descending order of stars), it became clear that most of these projects were not suitable for our study. In particular, out of these 1,000, we found only 40 projects to match all our selection criteria. Then, upon further inspection, other problems were found also for most of these 40 projects. Some of them, while having defined an explicit label for bugs, had very few labeled issues in the issue tracker. For others, while in the manual inspection of the change-log we observed commits linked to closed issues, the number of these links turned out to be very low even in projects having a very high number of commits and issues. This likely indicated the non-consistent adoption of a linking methodology between issues and commits. %Finally, several systems started using the issue tracker only recently, having a large part of their history without any information about issues.

For these reasons, we decided to adopt a different process for project selection. However, before describing it, we want to stress the challenges and perils of automatically selecting projects from GitHub. Indeed, while we applied some strong selection criteria on the number of issues (at least 100) and sorted projects based on their popularity as indicated by the number of stars (the most popular projects in our dataset had $\sim$67k stars), we obtained as result many tutorial-like projects (\eg~{\tt Snailclimb/JavaGuide}), repositories collecting quiz for job interviews (\eg~{\tt kdn251/interviews}) or, as previously said, repositories making a very limited use of methodologies to link commits and issues and/or to consistently label issues. We believe this is an important warning for our research community when dealing with large-scale studies in which project selection is not manually curated.
% TO BE REPORTED IN THE CONCLUSIONS

We decided to focus on projects managed by the Apache Software Foundation (ASF)~\cite{apache}, because these are well-used projects managed by a known open source foundation. Also, a large chunk of these projects consistently used through their entire change history a single bug-tracking system, namely JIRA~\cite{jira}. The issues are always classified based on their types (\eg bug) and, as a best practice, the Apache projects reference the issue id(s) in the note of commits closing issues. We used the GitHub API to extract the list of GitHub projects managed by the ASF. Then, we filtered out projects not having at least 80\% of their code written in Java, obtaining a list of 554 candidate projects. Finally, we sorted them by the number of forks (as a proxy for popularity), and two of the authors manually inspected this list from the top with the goal of selecting 100 projects to use for the study. The selection was done based on two criteria: 1) the project used the JIRA issue tracker for its entire change history; 2) the project was not a sub-project representing a ``component'' of a bigger project (\eg we excluded {\tt fineract-cn-portfolio}). If these two criteria were met, the authors annotated the name of the projects from the Apache JIRA installation \cite{jira} that were referenced in the change-log of the repositories (\ie in the commit notes). Indeed, the  Apache JIRA installation hosts several projects, each one identified by a specific name. For example, the {\tt apache/hadoop} project references in its change-log issues from the following projects hosted in Apache JIRA: HADOOP, HDFS, MAPREDUCE, and YARN. The two authors stopped when the set of 100 projects was collected (available in our online appendix~\cite{replication}).

For what concerns the three projects used in the replicated study, two of them (\ie ArgoUML and Xerces-J) use JIRA as well in their whole change history. Ant, instead, uses a mix of Bugzilla and JIRA and, thus, we had to manage this case in a different way as explained in the next section.

\subsection{Data Extraction}
\label{sec:data}
Once cloned the 103 repositories we used \tool{RMiner}~\cite{Tsantalis:icse2018} to identify commits containing refactoring operations. %Differently from \tool{Ref-Finder} 	\cite{PreteRSK10}, previously used by Bavota \etal \cite{BavotaCLPOS12}, \tool{RMiner} does not require to compile source code, and therefore it makes it possible to identify refactoring actions at commit level. Also, 
\tool{RMiner} has been estimated to achieve a precision of 98\% and a recall of 87\%. For each project, we run \tool{RMiner} on all commits of all branches impacting Java files, excluding merge commits. 

\tool{RMiner} outputs, for each commit, the list of refactoring actions detected, with the files and lines affected on the left-hand-side (before) and right-hand-side (after) of the change.

For the three projects studied by Bavota \etal \cite{BavotaCLPOS12}, we considered two different observation periods. The first considers the same history they analyzed \ie analyzing all commits preceding the releases they studied (identified from release tags or commit messages), and bug fixes limited within their observation period, \ie by December 31, 2011. Specifically, we considered the following release intervals: ArgoUML (0.11, 0.34], Ant (1.1, 1.8.2], and Xerces (1.0.3, 2.9.1].

The second observation period considers the whole evolution of the three projects up to January 15, 2020. Similarly, for the 100 Apache projects, we considered the entire history on GitHub until January 15, 2020.

To identify fix-inducing changes, we first download the issue reports of the mined projects by using the JIRA project names previously extracted during the project selection. For the 100 Apache projects, we download issue reports using the \tool{Perceval} tool \cite{perceval}. As for the three projects from the replicated study \cite{BavotaCLPOS12}, they use a heterogeneous way of reporting issues. While Xerces uses the Apache JIRA server, and  ArgoUML uses its own JIRA installation, Ant is the trickiest case because it used Bugzilla at the beginning of its history, and JIRA later. Also, Ant has several cases of bugs reported directly in the commit message. Therefore, for these projects, we identified regular expressions in commit messages referring to  (i) JIRA issues, (ii) Bugzilla issues, and (iii) bugs fixed without an issue. For the first two cases, we downloaded the issue reports using the \tool{wget} Unix utility, rendering them as free-text using the \tool{Lynx} browser, and extracted the relevant content using a Python script. For fixes without an issue report, we assumed the reporting and closing timestamp to match the commit timestamp.

Once downloaded the relevant issues, we linked them to commits using a regular expression-based approach~\cite{FischerPG03}. For the Apache projects, the regular expression is of type \texttt{ISSUEPROJECT-\#} (where ISSUEPROJECT is the name of the project on the issue tracker), whereas for the three other projects we used all possible regular expressions identified through the manual analysis explained above. We considered as bug-fixing commits those (i) linked to an issue of type ``Bug'' or, for Bugzilla (Ant), of priority at least ``Normal'' and not being an ``Enhancement''; (ii) where the issue was in status ``Closed'' and Resolution ``Fixed'', except for 12 Apache projects where the Closed status was not used, and we kept those with a ``Resolved'' status. For the Ant fixes without an issue, as explained before, we simply relied on the commit message regular expression. Finally, we noticed that some of the mined commits included commits reverting previous bug fixes (thus, they were matching our regular expressions since mentioning the issue for which they were reverting the fixing). We excluded these cases from the analysis.

While we are aware that software projects may contain fixes with no explicit link to issues~\cite{BirdBADBFD09} and that approaches to propose candidate links for such fixes exist~\cite{WuZKC11}, we preferred to avoid such a solution in order to limit false positives. 

More important, as explained in \secref{sec:context}, one criterion for the selection of projects was the careful usage of issue trackers (the only exception was Ant, which has several non-tracked issues, which we handled as explained above). We could have identified bugs from commits to mitigate the bias described by Bird \etal~\cite{BirdBADBFD09}, but this would have introduced false positives in the bug datasets and, also, would not have provided us with information about the issue opening date. For this reason, we limited this approach to untracked commits from Ant. %and while we select the 100 projects among those making  regular use of the issue tracker.

After having the set of bug-fixing commits and related issue metadata available, we were ready to apply the SZZ. 

At first, we tried to use already available tools, and in particular, \tool{SZZ Unleashed}~\cite{BorgSBH19}. However, by experimenting it and by discussing with its authors, we discovered that sometimes it tracks to wrong file version and line numbers, due to issues with the used Python git library. Thus, we implemented our own version of SZZ, capable of (i) ignoring cosmetic changes \new{(\ie formatting, using the \tool{git blame -w} option)}, changes to comments, and changes to non-Java files; and (ii) relying on the native Unix {\tt git diff}, renaming and line mapping. Our SZZ does not ignore semantically-equivalent changes because, indeed, we are interested in analyzing refactoring actions. Our SZZ implementation first identifies the lines changed by the fix. Then, starting from the file version before the fix, and considering only the fixed lines, it uses {\tt git blame -w -p} to identify the last change before the fix to these lines, along with the file name, and the line number mapping. In summary, for each changed line of fixed files, the algorithm outputs a candidate introduction location (commit, file name and line number). We discard candidate fix-inducing changes that occurred after the issue opening date. As for fixes without an issue (only for the Ant project), this heuristic was not used as a filter.

Recent work suggests that for an accurate fix-inducing change identification, bulk commits as well as the first commit of the project should also be ignored~\cite{CostaMSKCH17}, although the work also points out that such commits can still introduce fixes. For such reasons, we decided to keep them, also considering that (i) the first commit of the analyzed projects does not contain refactoring actions, and therefore false positives in those commits do not affect the experimental group; (ii) refactoring actions could occur in bulk commits because these can be commits aimed at performing a general restructuring of the projects. At the same time, in RQ$_2$ we control the effect of the change' size on the observed results. Furthermore, some SZZ implementations~\cite{DaviesRW14} only consider the most recent blame from each fix as a fix-inducing change, while we consider all possible blames as we want to be conservative. Indeed, we keep track of these changes and we show how results change if limiting the analysis only to those.

As a final step of our data extraction approach, we merge the SZZ output with the \tool{RMiner} output. Specifically, for each commit considered by \tool{RMiner}, we report:
\begin{enumerate}
\item whether it contains at least a refactoring;
\item whether it induces a fix;
\item whether there is at least one fix inducing change and refactoring action occurring in the same file;
\item whether there is at least one fix inducing change and refactoring action occurring on the same line;
\item detailed information for each refactoring action, \ie refactoring type and whether the refactoring occurs in a file and in a line with fix-inducing changes.
\end{enumerate}

Finally, to control for the size of the change, we compute using {\tt git diff}, for each analyzed commit, the number of changed Java files and the number of churns and of lines added and deleted in these files.

\subsection{Analysis Methodology}
\label{sec:methodology}

The analyses described below have been performed using the \tool{R} statistical environment~\cite{R}. To address \textbf{RQ$_1$}, we first use a methodology similar to the one applied by Bavota \etal~\cite{BavotaCLPOS12}. That is, we use Fisher's exact test~\cite{fisher} and Odds Ratio (OR) effect size to check whether commits containing at least one refactoring induce fixes in a higher proportion with respect to other commits. An OR $x>1$ indicates that the odds for refactoring-related commits to induce fixes are $x$ times greater than other commits. Note that for a refactoring-related commit we assume that the refactoring induces a fix if (1) at last a  refactoring occurs on the same file where the fix is induced; or (2) the refactoring impacts the same lines changed in the bug-fixing commit. We analyze results for both options (1) and (2). \new{For option (1) it is possible that refactoring and bug fixing occur in different lines of the same file. As it will be explained in \secref{sec:data}, since the relationship between the refactoring and the bug fix is determined using a re-implementation of the SZZ algorithm~\cite{Sliwerski:msr2005}, the fix must occur after the refactoring.}
We perform the analysis on each project separately, and then we adjust $p$-values using the Benjamini–Hochberg procedure~\cite{bh}.

To address \textbf{RQ$_2$}, we first identify the change size indicator to be used, by analyzing the presence of a correlation (using Spearman's rank correlation) between different size indicators. Then, we test the null hypotheses H$_{0r}$: refactoring-related commits do not have a significantly different size from other commits, and H$_{0f}$: fix-inducing commits do not have a significantly different size from other commits. We first test such null hypotheses using Wilcoxon rank-sum test~\cite{wilcoxon}. We then consider all possible combinations of the two factors (\eg fix-inducing and refactoring-related, fix-inducing but not refactoring related, \etc), using Kruskal-Wallis test~\cite{kruskal} followed by a Dunn post hoc analysis~\cite{dunn}. We also report Cliff's delta effect size values~\cite{Cliff:2005}.

Finally, we study whether the size of the change and refactoring actions interact with respect to inducing a fix, by using a logistic regression model with mixed-effect (\emph{glmer} function of the \tool{R} \emph{lme4} package~\cite{lme4}). The dependent variable is a dichotomous variable indicating if a commit is fix-inducing or not; the independent variables are dichotomous variables indicating whether a commit contains at least a refactoring which impacted a fix inducing file or line, the commit size, and their interaction. The random effect is the project.

To address \textbf{RQ$_3$}, we perform, on data from all projects, an analysis similar to the one of RQ$_1$, but by refactoring type. That is, we consider whether commits containing at least one refactoring of a given type have higher odds to induce a fix (again considering as positive cases when the refactoring overlaps with the fix at file or line level) than commits not containing that kind of refactoring. Since the test is repeated for 41 refactoring types, $p$-values are adjusted as before.

To address \textbf{RQ$_4$}, we firstly extracted from our dataset the 17,985 bug-fixing commits for which a match with one or more refactoring was found at line level in the fix-inducing commit. This means that the source code lines impacted by the bug-fixing commit were also impacted, completely or in part, by refactoring operations performed in the fix-inducing commit. 

Once obtained this set, we extracted from it a statistically significant sample ensuring a 95\% confidence level $\pm$ 5\%. 

This resulted in the selection of 384 bug-fixing commits with their related refactoring operations. The selection of the 384 instances was performed in the following way. First, we analyzed the distribution of refactoring types (\eg \emph{extract class}, \emph{extract method}, \etc) in the entire population of fix-inducing commits implementing refactoring actions. In this way, we found out the percentage of fix-inducing commits in which each refactoring type appears. %For example, the most popular refactoring in our fix-inducing dataset was the \emph{extract method}, with 18.3\% of commits implementing one of these operations, followed by the \emph{change variable type} with 8.5\% \etc 
Then, we also computed the number of fix-inducing commits in each of the 103 systems considered in our study. The system and the refactoring type were used as strata to randomly select the 384 commits for manual validation. This means that the higher the number of fix-inducing commits in a system $S$, the higher the number of fix-inducing commits from $S$ that will be in our sample.  Similarly, the higher the number of fix-inducing commits containing a certain refactoring type $T$, the higher the number of commits implementing $T$ in our sample.

%TODO when unblinding ``by three'' ->''by the three''
The selected sample was manually analyzed by three authors (from now on evaluators) with the goal of classifying them as false positive (\ie the refactoring in the fix-inducing commit was not responsible for the bug introduction) or as a true positive (\ie the refactoring introduced the bug). In the latter case, the evaluator could also briefly describe the reason why the refactoring induced the bug-fixing activity.

The manual analysis was supported by a Web app that we developed for this task. Each author independently inspected the commits randomly assigned to her by the Web app. Each commit was assigned to two evaluators by the Web app, that showed for a given commit: (i) the link to the bug-fixing commit in GitHub, highlighting the code line(s) modified in it that was also impacted by the refactoring; (ii) the link to the fix-inducing commit in GitHub, highlighting the code line(s) impacted by the refactoring that was also modified in the bug-fix; (iii) a list of the refactoring actions detected by \tool{RMiner} that were implemented in the fix-inducing commit and matched the lines in the bug-fix. Each author roughly classified 270 commits to obtain the two evaluations needed for each of the 384 commits. At the end of this process, the authors performed an open discussion to solve the 117 conflicts (30\%) that have occurred. 

To answer RQ$_4$, we report the percentage of analyzed commits in which we found an actual link between refactoring and bug introduction. Also, we discuss interesting cases identified in our manual analysis. 
%%%%%%%%%%%%%%%%%%%%%%%%%%%%%%%%
%%%%%%%%%%%%%%%%%%%%%%%%%%%%%%%%
\section{Study Results}
\label{sec:results}
%%%%%%%%%%%%%%%%%%%%%%%%%%%%%%%%
%%%%%%%%%%%%%%%%%%%%%%%%%%%%%%%%

We discuss the results accordingly to the defined RQs.\smallskip

\textbf{RQ$_1$: \rqone} We report the comparison of the proportion of fix-inducing changes occurring in commits with a refactoring --- overlapping at the file(s) or the lines(s)  level --- and in other changes. In particular, \tabref{tab:rq1:samehistory} reports results on the same systems and on the same history studied by Bavota \etal~\cite{BavotaCLPOS12}. 

As the table shows, commits with refactoring always have significantly higher odds to induce a fix than other changes. Looking at the top-side of the table, the ORs are between 3.46 and 3.87 when considering a matching at the file level. This is the closest comparison to  Bavota \etal~\cite{BavotaCLPOS12}: compared to our results, they reported OR at release-level, with the following OR ranges computed for significant differences: Ant [3.50,6.65], ArgoUML 5.17 (only one release showed significant results) and between 8.79 and 157.69 for Xerces2-J. Note that they counted proportions on refactoring instances detected on a single release, and because of that for many releases (13 out of 17 for Ant, 13 out of 14 for ArgoUML, 23 out of 29 for Xerces2-J) they did not obtain statistically significant results. However, such a lack of significance seems to be due more to a limited statistical power rather than to other reasons. Similarly, the 157 OR they observed for ArgoUML was computed on a release with only 2 refactoring actions and 2 fix inducing commits.

\begin{table}[t]
	\caption{RQ$_1$: Replication on the same systems and history of Bavota \etal. (NRNI: no refactoring, no inducing fix; NRI: no refactoring, inducing fix; RNI: refactoring, no inducing fix; RI: refactoring, inducing fix).}
	\label{tab:rq1:samehistory}
	\centering
	\begin{tabular}{lrrrrrr}
		\hline
		\multicolumn{7}{c}{\textsc{File matching}}\\
		\hline
			\textbf{System} & \textbf{NRNI} & \textbf{NRI} & \textbf{RNI} & \textbf{RI} & \textbf{OR} & \textbf{p adj} \\ 
		\hline
		Ant & 11,823 & 288 & 1,981 & 187 & 3.87 & \textbf{$<$0.001} \\ 
		ArgoUML & 19,413 & 458 & 3,979 & 344 & 3.66 & \textbf{$<$0.001} \\ 
		Xerces2-J & 4,206 & 614 & 672 & 340 & 3.46 & \textbf{$<$0.001} \\ 
		\hline
		\multicolumn{7}{c}{\textsc{Line matching}}\\
		\hline
			\textbf{System} & \textbf{NRNI} & \textbf{NRI} & \textbf{RNI} & \textbf{RI} & \textbf{OR} & \textbf{p adj} \\ 
		\hline
		Ant & 11,823 & 288 & 2,085 &  83 & 1.63 & \textbf{$<$0.001}  \\ 
		ArgoUML & 19,413 & 458 & 4,144 & 179 & 1.83 & \textbf{$<$0.001}  \\ 
		Xerces2-J & 4,206 & 614 & 846 & 166 & 1.34 & \textbf{$<$0.001}  \\ 
		\hline
	\end{tabular}
\end{table}

Looking at the bottom side of the table, if we consider that a refactoring induces a fix only if a line affected by the refactoring is also modified in the bug-fixing commit, odds are reduced by 60\% or more, and vary between 1.34 and 1.83. Still, changes involving refactoring actions have higher odds to induce a fix. Also, note this is a very conservative analysis because a refactoring might still impact a fix without directly affecting a line modified in the bug fix.

\begin{table}[t]
	\caption{RQ$_1$: Replication on the same systems of Bavota \etal, history up to date. (NRNI: no refactoring, no inducing fix; NRI: no refactoring, inducing fix; RNI: refactoring, no inducing fix; RI: refactoring, inducing fix).}
	\label{tab:rq1:samesystems}
	\centering
	\begin{tabular}{lrrrrrr}
		\hline
		\multicolumn{7}{c}{\textsc{File matching}}\\
		\hline
		\textbf{System} & \textbf{NRNI} & \textbf{NRI} & \textbf{RNI} & \textbf{RI} & \textbf{OR} & \textbf{p adj} \\ 
		\hline
		Ant & 13,762 & 371 & 2,248 & 257 & 4.24 & \textbf{$<$0.001} \\ 
		ArgoUML & 22,526 & 590 & 4,366 & 410 & 3.59 & \textbf{$<$0.001} \\ 
		Xerces2-J & 5,747 & 686 & 970 & 392 & 3.38 & \textbf{$<$0.001} \\ 
		\hline
		\multicolumn{7}{c}{\textsc{Line matching}}\\
		\hline
			\textbf{System} & \textbf{NRNI} & \textbf{NRI} & \textbf{RNI} & \textbf{RI} & \textbf{OR} & \textbf{p adj} \\  
		\hline
		Ant & 13,762 & 371 & 2,384 & 121 & 1.88 & \textbf{$<$0.001} \\ 
		ArgoUML & 22,526 & 590 & 4,564 & 212 & 1.77 & \textbf{$<$0.001} \\ 
		Xerces2-J & 5,747 & 686 & 1,168 & 194 & 1.39 & \textbf{$<$0.001} \\ 
		\hline
	\end{tabular}
\end{table}

Considering the complete history of the projects, as \tabref{tab:rq1:samesystems} shows, results are quite consistent with the ones of \tabref{tab:rq1:samehistory}.

\begin{figure}[tb]
	\centering
	\includegraphics[width=0.9\linewidth]{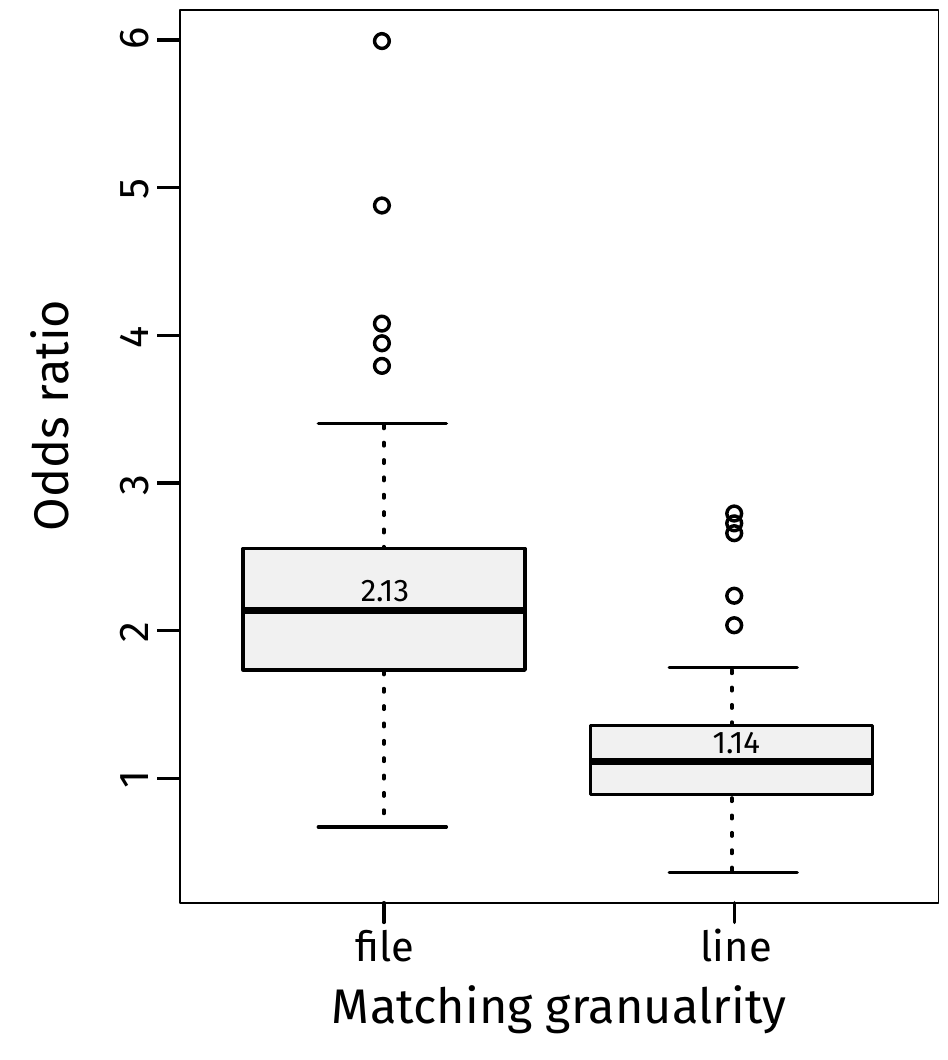}
	\caption{RQ$_1$: Odds that refactoring actions induce fixes. Boxplot of OR for the 100 Apache projects.}
	\label{fig:rq1-or}
\end{figure}

When performing the Fisher's exact test for the 100 Apache projects, at file-level, 85 $p$-values are statistically significant ($<0.05$, before and after the adjustment). At line-level, only 34 $p$-values are statistically significant, 28 after the adjustment. \figref{fig:rq1-or} shows the distribution of OR at file- and line-level matching for the 100 Apache projects. An OR greater than one indicates that a commit where a refactoring occurs has more chances than other commits to induce a fix. For the file-level matching, the median OR is 2.13 (it reaches 2.36 if considering only the projects where the difference in proportion is statistically significant). For the line-level matching, the OR decreases dramatically to a level at which the difference between a commit with refactoring actions and other commits is smaller (OR=1.13, while it reaches 1.46 if considering statistically significant cases only).

What if considering as fix-inducing only the most recent blame~\cite{DaviesRW14}? We performed the analysis (details in the replication package~\cite{replication}), and results did not change dramatically. For the three projects of Bavota \etal, odds were still above 3 at file-level and above 1.5 at line-level. For the 100 Apache projects the median OR was 2.07 and 1.17 at file- and line-level, respectively. 

\begin{resultbox}
	\textbf{RQ$_1$ Summary:} Our results confirm the main findings of Bavota \etal~\cite{BavotaCLPOS12}. Commits implementing refactoring actions have higher odds to induce a fix than other changes. This finding is also confirmed when working at line-level granularity, even though the difference between refactoring and other types of changes is less marked.
\end{resultbox}

\textbf{RQ$_2$: \rqtwo}  We found a moderate to strong correlation (0.59) between the number of changed files and the number of added lines, between the number of added and deleted lines (0.46), between the number of added lines and added churns (0.79), and between the number of added lines and deleted churns (0.50). Therefore, we only report our analysis considering, as the size of a change, the number of added lines. We also performed the same analysis for the other factors, obtaining similar results.

%\begin{figure}[tb]
%	\centering
%	\includegraphics[width=\linewidth]{img/RQ2}
%	\caption{RQ$_2$: Relationship between fix-inducing changes and refactoring commits with the size of a change: boxplots and interaction plots. \MAX{figure to be improved!}}
%	\label{fig:rq2}
%\end{figure}

%The Wilcoxon rank sum test indicates that commits with fix-inducing changes are bigger than other commits ($p$-value$<$0.001) with a medium effect size ($d$=0.40), see also the boxplots in the top-right corner of \figref{fig:rq2}. At the same time, commits in which refactoring occur introduce significantly bigger than others ($p$-value$<$0.001), with a large effect size ($d$=0.50), see also the boxplot in the bottom-left corner of \figref{fig:rq2}. 

%The interaction plots (top-left and bottom-right corners of  \figref{fig:rq2})  show that while both factors have an effect on the size, the interaction is mild. Since on our dataset the conditions for ANOVA application were not met (residuals not normally distributed and variance not homogeneous), we verified the presence of interaction using a Kruskal-Wallis test followed by a post-hoc Dunn's test \cite{dunn} with Benjamini-Hochberg correction. The test indicates that all possible combinations are statistically different from each other, and that (i) changes with refactorings and fix-inducing changes are larger than all other changes, changes with refactorings but not fix-inducing are larger than changes with no refactoring but fix inducing, and changes with no refactorings and not fix-inducing are smaller than any other group.

The Wilcoxon rank-sum test indicates that commits with fix-inducing changes are bigger than other commits ($p$-value $<$ 0.001) with a medium effect size ($d$=0.40). At the same time, commits in which refactoring actions occur are significantly bigger than others ($p$-value $<$ 0.001), with a large effect size ($d$=0.50). In our dataset the conditions for ANOVA application were not met (residuals not normally distributed and variance not homogeneous). Therefore, we verified the presence of interaction between the two factors (\ie refactoring and fix-inducing) using a Kruskal-Wallis test followed by a post hoc Dunn's test with Benjamini-Hochberg correction. The test indicates that all possible combinations are statistically different from each other, and that (i) changes with refactoring actions and fix-inducing changes are larger than all other changes; (ii) changes with refactoring actions but not fix-inducing are larger than changes with no refactoring but fix-inducing, and (iii) changes with no refactoring actions and no fix-inducing are smaller than any other group.

\begin{table}[t]
\caption{Mixed-effect logistic regression relating refactoring, lines added, and their interaction with fix-inducing changes}
\label{tab:rq2:logistic}
	\centering
	\small
	\resizebox{1\linewidth}{!}{
	\begin{tabular}{rrrrr}
		\hline
		 \textbf{AIC}    &     \textbf{BIC}  &   \textbf{logLik}  &  \textbf{deviance}  &  \textbf{df.residuals} \\
		300,772.4   & 300,828.6  & -150,381.2  &  300,762.4   &   562,671  \\
		\hline
		\multicolumn{5}{l}{\textsc{Scaled residuals:}}\\
		\hline
		\textbf{Min}    &  \textbf{1Q} & \textbf{Median } &    \textbf{3Q}  &   \textbf{Max} \\ 
		-3.4245  & -0.3249  & -0.2389  & -0.1491  & 15.0509 \\
		\hline
		\multicolumn{5}{l}{\textsc{Random effects:}} \\
		\hline
		\multicolumn{3}{l}{\textbf{Groups}}  &      \textbf{Variance} & \textbf{Std.Dev.}  \\
		\multicolumn{3}{l}{Project (Intercept)} & 0.7136 &  0.8448 \\ 
		\multicolumn{5}{l}{Number of obs: 562,676, Groups:  Project: 103}\\
		\hline
		\multicolumn{5}{l}{\textsc{Fixed effects:}}\\
		\hline
		& \textbf{Estimate} & \textbf{Std. Error} & \textbf{z value} & \textbf{Pr($>$$|$z$|$)} \\ 
		\hline
		(Intercept) & -3.24 & 0.08 & -39.81 & \textbf{$<$0.001} \\ 
		Ref. & 0.83 & 0.01 & 61.97 & \textbf{$<$0.001} \\ 
		Lines Added & 0.01 & 0.00 & 80.96 & \textbf{$<$0.001} \\ 
		Ref.:Lines Added & -0.00 & 0.00 & -16.95 & \textbf{$<$0.001}\\ 
		\hline
	\end{tabular}}
\end{table}

Finally, we use a mixed-effect logistic regression model to evaluate whether, even in presence of the ``size'' effect, refactoring actions still correlate with fix-inducing changes. As \tabref{tab:rq2:logistic} shows, the occurrence of refactoring actions, the commit size in lines added, and their interaction have a statistically significant effect on the likelihood that the commit induces a fix. By observing the estimates, the presence of a refactoring increases by $e^{0.83}=2.29$ times the odds that a commit induces a fix, while a unity increment of the added lines increases the odds by $e^{0.01}=1.01$, and a similar effect size is observed for the interaction between refactoring actions and lines added. 

Similar results have been obtained considering, as a change size indicator, the number of added churns.

\begin{resultbox}
	\textbf{RQ$_2$ Summary:} When controlling for size, the refactoring actions still play a role in inducing bug-fixing activities, thus supporting the RQ$_1$ findings.
\end{resultbox}

\begin{table}[t]
	\caption{Fix-inducing proneness by type of refactoring.}
	\label{tab:rq3:type}
	\centering
	\small
	\resizebox{1\linewidth}{!}{
	\begin{tabular}{lrrrrr}
		\hline
		\textbf{Name} & \textbf{\#} & \textbf{(\%)} & \textbf{Buggy} & \textbf{OR} & \textbf{p adj} \\ 
		\hline
		Extract Subclass & 910 & 0.31 & 232 & 2.07 & \textbf{$<$0.001} \\ 
		Move And Inline Method & 1,962 & 0.68 & 422 & 1.65 & \textbf{$<$0.001} \\ 
		Extract Class & 4629 & 1.60 & 994 & 1.65 & \textbf{$<$0.001} \\ 
		Extract And Move Method & 6,633 & 2.29 & 1,331 & 1.52 & \textbf{$<$0.001} \\ 
		Move And Rename Method & 3,672 & 1.27 & 735 & 1.51 & \textbf{$<$0.001} \\ 
		Push Down Method & 744 & 0.26 & 143 & 1.44 & \textbf{$<$0.001} \\ 
		Split Attribute & 413 & 0.14 &  73 & 1.30 & 0.07 \\ 
		Extract Superclass & 5,272 & 1.82 & 916 & 1.27 & \textbf{$<$0.001} \\ 
		Merge Variable & 838 & 0.29 & 140 & 1.21 & 0.06 \\ 
		Move Method & 5,767 & 1.99 & 930 & 1.15 & \textbf{$<$0.001} \\ 
		Parameterize Variable & 3,870 & 1.33 & 618 & 1.15 & \textbf{$<$0.001} \\ 
		Merge Parameter & 738 & 0.25 & 117 & 1.14 & 0.25 \\ 
		Replace Attribute & 203 & 0.07 &  32 & 1.13 & 0.52 \\ 
		Split Parameter & 331 & 0.11 &  52 & 1.13 & 0.48 \\ 
		Extract Interface & 2,011 & 0.69 & 312 & 1.11 & 0.14 \\ 
		Split Variable & 149 & 0.05 &  23 & 1.10 & 0.65 \\ 
		Inline Method & 5,056 & 1.74 & 782 & 1.10 & \textbf{0.03} \\ 
		Push Down Attribute & 575 & 0.20 &  88 & 1.09 & 0.48 \\ 
		Pull Up Attribute & 1,310 & 0.45 & 198 & 1.08 & 0.42 \\ 
		Pull Up Method & 1,571 & 0.54 & 230 & 1.03 & 0.65 \\ 
		Move Attribute & 4,614 & 1.59 & 656 & 1.00 & 1.00 \\ 
		Move And Rename Class & 2,999 & 1.03 & 389 & 0.90 & 0.09 \\ 
		Replace Variable With Attr. & 3,469 & 1.20 & 443 & 0.88 & \textbf{0.02} \\ 
		Extract Method & 27,371 & 9.44 & 3,509 & 0.86 & \textbf{$<$0.001} \\ 
		Merge Attribute & 576 & 0.20 &  70 & 0.84 & 0.22 \\ 
		Move And Rename Attr. & 216 & 0.07 &  25 & 0.79 & 0.40 \\ 
		Inline Variable & 5,957 & 2.05 & 611 & 0.69 & \textbf{$<$0.001} \\ 
		Rename Attribute & 15,435 & 5.32 & 1,535 & 0.66 & \textbf{$<$0.001} \\ 
		Rename Variable & 23,327 & 8.05 & 2,310 & 0.65 & \textbf{$<$0.001} \\ 
		Change Parameter Type & 15,098 & 5.21 & 1,446 & 0.63 & \textbf{$<$0.001} \\ 
		Change Variable Type & 20,484 & 7.07 & 1,929 & 0.61 & \textbf{$<$0.001} \\ 
		Rename Parameter & 19,011 & 6.56 & 1764 & 0.60 & \textbf{$<$0.001} \\ 
		Change Return Type & 16,868 & 5.82 & 1,493 & 0.58 & \textbf{$<$0.001} \\ 
		Change Attribute Type & 20,064 & 6.92 & 1,721 & 0.56 & \textbf{$<$0.001} \\ 
		Rename Method & 20,938 & 7.22 & 1,798 & 0.54 & \textbf{$<$0.001} \\ 
		Extract Variable & 25,328 & 8.74 & 2,150 & 0.54 & \textbf{$<$0.001} \\ 
		Rename Class & 8,459 & 2.92 & 609 & 0.46 & \textbf{$<$0.001} \\ 
		Extract Attribute & 2785 & 0.96 & 197 & 0.46 & \textbf{$<$0.001} \\ 
		Move Class & 6,345 & 2.19 & 373 & 0.37 & \textbf{$<$0.001} \\ 
		Change Package & 1,149 & 0.40 &  60 & 0.33 & \textbf{$<$0.001} \\ 
		Move Source Folder & 2,750 & 0.95 &  58 & 0.13 & \textbf{$<$0.001} \\ 
		\hline
	\end{tabular}}
\end{table}

\textbf{RQ$_3$: \rqthree} \tabref{tab:rq3:type} reports, for each refactoring type, the odd that a commit containing at least a refactoring of that type has to induce a fix. For this RQ, for space reasons, we consider only the case in which the refactoring overlaps with the bug-fix at line level. This also because non-overlapping lines in the same file could be subject to other refactoring types. Refactoring types are ordered by decreasing OR.

Most of the refactoring types having a high odd to induce fixes are those involving refactoring big chunks of code (\emph{extract class}/\emph{subclass}, \emph{move and inline method}/\emph{extract and move methods}), as well as those involving inheritance (\emph{extract subclass}/\emph{superclass}, \emph{push down method}). 

The latter confirms previous findings~\cite{BavotaCLPOS12}, which also found such types of refactoring to be particularly concerning, and literature highlighting the difficulties to test class hierarchies~\cite{HarroldMF92}. %\MAX{more refs on inheritances and fault proneness}.

We can also notice how some refactoring actions not involving large changes, \eg \emph{split attribute} and \emph{merge variable} have a relatively high OR (1.30 and 1.20, respectively). Instead, renaming changes are largely harmless, despite being among the most frequent refactoring actions we found. Surprisingly, \emph{extract method}, another very frequent refactoring has an OR (0.86) smaller than similar refactoring types (\eg \emph{extract and move method}, 1.52). It is possible that extracting a method within the same class creates fewer problems than an \emph{extract and move method} (due to the need for context adjustment).

\begin{resultbox}
	\textbf{RQ$_3$ Summary:} Twenty refactoring types confirm their higher chances to induce fixes as compared to other types of changes, with ten of them being statistically significant. As compared to the work by Bavota \etal~\cite{BavotaCLPOS12}, we confirm the high odds to induce fixes for refactoring types related to inheritance.
\end{resultbox}

\begin{figure}[tb]
	\centering
	\includegraphics[width=0.85\linewidth]{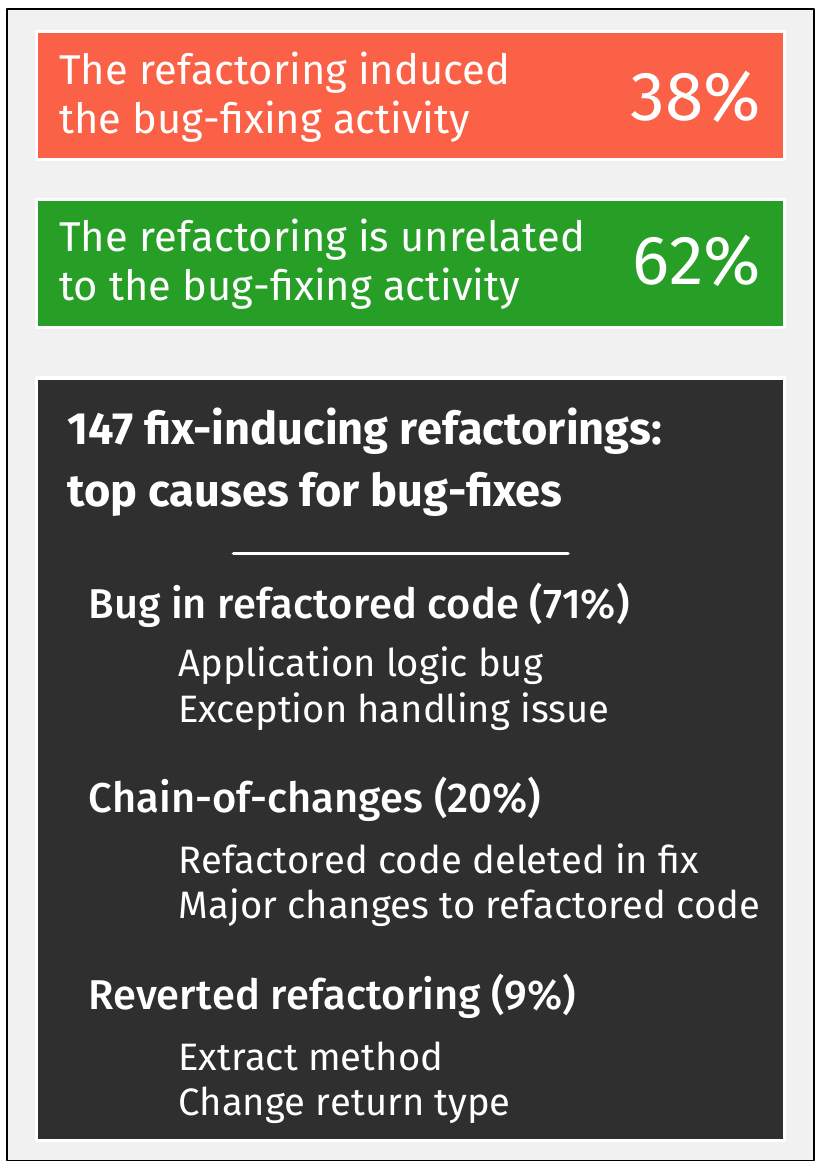}
	\caption{RQ$_4$: Manual validation of 384 fix-inducing commits implementing refactoring actions.}
	\label{fig:rq4}
\end{figure}

\subsection{RQ$_4$: \rqfour}
\figref{fig:rq4} shows the results of the manual validation we performed to verify whether the refactoring actions detected in 384 fix-inducing commits identified through the SZZ algorithm were actually responsible for triggering the bug-fixing activity. Before commenting on the results, a number of clarifications must be made. First, we noticed that in some cases what was labeled as ``bug'' in the issue tracker of the subject systems was not a \emph{functional} bug, but rather an issue with non-functional aspects of source code (\eg performance) or, in a few cases, minor issues (\eg a wrong logging message). 

We do not make any distinction among these types of issues in our study, assuming that what was labeled by the original developers as a ``bug'' should be considered as such. %Thus, if a refactoring was responsible for introducing a performance issue and such an issue was labeled as a bug, we consider the refactoring as triggering a bug. 
Second, while the authors involved in the manual validation have a strong experience in Java (\ie the language used in all subject systems), they are not the developers of the subject systems. In some cases, while we managed to identify the refactored code as responsible (or not) for triggering the fixing activity, we found extremely difficult to distill the exact code change that caused the bug. For example, let us assume that a method created through an \emph{extract method} refactoring in the fix-inducing commit was the target of changes in the bug-fix, and that the impacted code was created during the \emph{extract method} refactoring (\ie did not previously exist in the system). We labeled this refactoring as fix-inducing even if we did not manage to locate the actual bug in the code.

For 147 (38\%) of the analyzed fix-inducing commits, we classified the refactoring as responsible for triggering the bug-fixing activity. This means that in 62\% of cases (237 commits), while the refactoring actions were part of the changes implemented in the fix-inducing commit, the manual analysis did not show any evidence about their implication in the bug introduction. The main reasons for not considering the refactoring as the trigger for the fixing commit were three. In 31\% of cases (74), the refactored code was unrelated to the bug introduction meaning that, while the bug was actually introduced in the commit indicated by the SZZ (\ie the one implementing the refactoring) and the bug-fixing commit also modified lines of code impacted by the refactoring, the fixed bug concerned other lines modified in the same commit that were not subject of any refactoring activity. In 29\% of cases (68 out of 237), the fixed bug already affected the system before the refactoring. An example of this scenario is the case in which an \emph{extract class} refactoring grouped together a number of existing statements and one of them was already buggy (\eg the condition in an {\tt if} statement, then fixed in the bug-fixing commit). The subsequent \emph{extract class} did not change the statements but was identified by the SZZ as responsible for triggering the fix since it was the last change impacting on the buggy statement. Finally, in the remaining 40\% of cases (95), the refactoring and or the bug-fixing were part of tangled commits \new{(such a percentage is smaller of the proportion of floss refactoring indicated in previous literature~\cite{Emerson:tse2011}, \ie about 60\%, but not particularly small)}, often of huge size, that made extremely difficult to identify the actual triggering of the bug-fix. However, in all those cases, the authors agreed on the unlikely link between the refactoring and the bug introduction.

For what concerns the 147 ``true positive'' instances, \figref{fig:rq4} shows the three causes we identified for the triggering of bug-fixing activities, \ie \emph{Bug in refactored code}, \emph{Chain-of-changes}, and \emph{Reverted refactoring}. Each of these categories contains sub-categories better detailing the reason behind the bug. Due to space limitations, we only report in \figref{fig:rq4} the top-2 subcategories for each of these main categories. The complete categorization is available in our online appendix~\cite{replication}. In the following, we describe each category and present one representative example for each of them.

\emph{\textbf{Bug in refactored code}}. This is the ``obvious'' and expected reason for which a refactoring should trigger a bug-fixing commit and, indeed, this category accounts for 71\% of the true positive cases. Most of the bugs in this category are related to application logic bugs, to the handling of exceptions, and to wrong initialization of variables. An example of this category is commit {\tt c20ac05} from {\tt apache/karaf}, in which an \emph{extract method} refactoring is implemented. In particular, part of the {\tt doExecute} method from the {\tt DisplayLog} class is extracted into the newly created {\tt display} method, which is then invoked in {\tt doExecute} through the statement {\tt display(cnv,\- event,\- out)}. The bug-fixing commit ({\tt d9ecb3d}), which commit note mentions ``\emph{[KARAF-546] Added NPE check inside DisplayLog}'', adds a Null Pointer Exception (NPE) guard in an {\tt if} statement preceding the invocation of the extracted method (\ie {\tt if(event != null)}) and avoids possible NPE. The changes introduced due to the performed \emph{extract method} have induced the bug-fixing commit. 

\emph{\textbf{Chain-of-changes}}. In 20\% of cases, while we were not able to precisely identify bugs in the refactored code, we observed a ``chain-of-changes'' triggered by the refactoring and resulting in the bug-fixing commit. For example, in 10\% of cases, the refactored code (\eg an extracted variable/class/method) was deleted in the bug-fixing commit. In the remaining cases, the bug-fixing commit implemented major changes in the previously refactored code. For example, we found seven commits in which the refactoring changed the type of a parameter, a variable, or the return type of a method and then, the bug-fixing commit changed that same type again --- not to the original type (\ie the one before the refactoring) but to a new one. An example of these cases is commit {\tt 6a1ced0} from {\tt apache/felix}, in which the developers performed a \emph{change parameter type} refactoring, changing {\tt Source sourceDirectory} to {\tt List sourceDirectories}. The bug-fixing commit changed again the parameter to {\tt File outputDirectory}, with a consequent impact on the application logic of the method. Note that in these cases the link between refactoring and bug-introduction is less strong as compared to the previous category. However, we still see the refactoring as at least one of the causes of the changes implemented in the bug-fix.

\emph{\textbf{Reverted refactoring}}. In 9\% of cases, the bug-fixing commit reverted the changes implemented by the refactoring. Differently from the \emph{Chain-of-changes} category, in this case the refactored code was reverted to its status before the refactoring. The most reverted refactoring actions are those related to the changes of types. In commit {\tt ae008b7} of {\tt apache/hive}, a \emph{change variable type} converts the type of a variable {\tt t} from {\tt TimestampWritable} to {\tt TimestampWritableV2}. Such a change is reverted in the bug-fixing commit {\tt bd95a2f} with the following comment added in the source code {\tt //Use old\- timestamp\- writable\- hash- code\- for\- backwards\- compatibility}. 

\new{It is important to note that, while we found 9\% of reverted refactoring, none of them belong to an explicitly reverted commit (previous research indicate how reverted commits are used to undo changes throughout a project's history \cite{ShimagakiKMPU16}.)}

\begin{resultbox}
	\textbf{RQ$_4$ Summary:} Our manual validation, while confirming the possible role played by refactoring in the introduction of bugs, partially debunks the findings of our quantitative analysis and of previous studies~\cite{BavotaCLPOS12}. Indeed, in 62\% of cases, while the SZZ reports the commit implementing refactoring as the one inducing the bug-fixing activity, we did not find evidence of the linking between the refactored code and the bug-fix.
\end{resultbox}
%%%%%%%%%%%%%%%%%%%%%%%%%%%%%%%%
%%%%%%%%%%%%%%%%%%%%%%%%%%%%%%%%
\section{Threats to Validity}
\label{sec:threats}

\emph{Construct validity.} Imprecisions in the detected refactorings could have affected our results. However, we used a highly precise state-of-the-art tool (\tool{RMiner}~\cite{Tsantalis:icse2018}), reported to have a 98\% precision and 87\% recall. Another threat is related to the approximations and the granularity of the SZZ algorithm~\cite{BorgSBH19} used for identifying fix-inducing changes. As detailed in \secref{sec:data}, we used appropriate heuristics to mitigate this issue, \eg filter out commented code and cosmetic changes.
\new{Although we did not compute the accuracy for our SZZ re-implementation, we mitigate this threat (i) by testing our implementation on a set of $\sim$ 20 bug introduction instances, and (ii) through the manual analysis performed in the context of \RQ{4}.}

Finally, links between commits and issues may be missing and biased~\cite{BirdBADBFD09}, or issues improperly tagged~\cite{AntoniolAPKG08,HerzigJZ15}. This is one of the reasons why we decided to use as subject systems a set of projects adopting well-defined practices to label issues and to link them to commits.

\emph{Conclusion validity.} As already detailed in \secref{sec:methodology}, wherever possible we used appropriate statistical procedures with $p$-value correction and effect size measures to test the significance of the differences and their magnitude. 

\emph{Internal validity.} Those are mainly related to a missing causation link between refactorings and bug fixes and to possible confounding factors that may influence such a relationship. We controlled for the size of implemented changes as confounding factors.  Other co-factors not considered in our study may play a role in the reported findings (\eg floss refactoring activities). However, (i) in our observational study we do not claim causation, and (ii) at least, we complemented the quantitative analysis with a  qualitative one, which helped in better understanding the refactoring-bug relationship. 

\emph{External validity.} While we considered over 100 projects in our study, we only considered Java projects belonging to the Apache ecosystem. \new{In \secref{sec:context}, we explained the reasons of this choice, \ie availability of reliable-enough defect data.} Our findings may not generalize to other languages or to systems outside of this ecosystem. Also, we only considered the refactoring operations currently supported by \tool{RMiner}.

%%%%%%%%%%%%%%%%%%%%%%%%%%%%%%%%
%%%%%%%%%%%%%%%%%%%%%%%%%%%%%%%%
\section{Related Work}
\label{sec:related}
%%%%%%%%%%%%%%%%%%%%%%%%%%%%%%%%
%%%%%%%%%%%%%%%%%%%%%%%%%%%%%%%%

As reported in the introduction, many studies have investigated software refactoring practices \cite{Emerson:tse2011,Kim:fse2012,Silva:fse2016,Peruma:iwor2018}. In this section, we focus on the ones aimed at investigating the impact of refactoring on code quality, since being the most related to our work.
%\MAX{should we discuss better Kim:fse2012? although we already did in the intro...}

%\subsection{On Refactoring and Code Quality}
Bavota \etal~\cite{Bavota:jss2015}, mined the evolution history of three open source projects looking at whether refactoring operations usually involve code components with specific characteristics in terms of quality metrics and presence of smells. 

Their results highlight that (i) very often quality metrics do not show a clear relationship with refactoring; (ii) only 42\% of refactoring involves code components affected by code smells; and (iii) only 7\% of the performed operations actually remove the code smells from the affected class.

Cedrim \etal~\cite{Cedrim:fse2017} conducted a longitudinal study aimed at characterizing the beneficial and harmful effects of refactoring on code smells. %They analyzed how frequently refactoring operations affect the density of 13 code smells by looking at the evolutionary histories of 23 projects. 
Their results show that even if in $\simeq 80\%$ of cases refactoring activities involve smelly elements, only $\simeq 10\%$ of the refactoring actions results in the removal of code smells from the affected code. Moreover, they found that while applying refactoring developers tend to introduce new code smells (33\%), \eg $\simeq 30\%$ of \emph{move method} and \emph{pull up method} refactoring operations introduce a God Class.

Ch{\'a}vez \etal~\cite{chavez2017does} analyzed the impact of refactoring on internal quality attributes by looking at 29k refactoring actions occurred in the history of 23 projects. They found that often the refactoring touches code components showing at least one critical internal quality attribute. Furthermore, they show that 55\% of these operations improve internal quality attributes against a 10\% of code quality decline.

Eposhi \etal~\cite{eposhi2019removal} studied, among other things, the relationship between refactoring and code quality issues. Their findings show that (i) the density of code smells is more than 8 times higher in refactored classes and (ii) refactoring actions usually do not reduce the density of quality issues.

Bibiano \etal~\cite{bibiano2019quantitative} looked at refactoring operations applied in batches rather than in isolation to analyze their effect on code smells. Their study is based on the assumption that a single refactoring rarely suffices to remove a code smell. Surprisingly, their results show that batches mostly ended up introducing (51\%) or not fully removing (38\%) smells.

Vassallo \etal~\cite{Vassallo:scp2019} mined 200 systems to quantitatively investigate factors correlating with refactoring, looking at when, why, and by whom refactoring is performed. Their results show that refactorings (i) are rarely performed close to a new release; (ii) are mainly performed while improving existing features; and (iii) are mainly done by the owners of the code components being refactored.

All the aforementioned work relate refactoring actions to quality attributes, such as metrics, code smells, or to process indicators (as Vassallo \etal~\cite{Vassallo:scp2019} did), whereas our study relates refactoring actions to bug introduction, while considering the effect of some change metrics (\ie change size) as a co-factor. Our study allowed to (partially) corroborate previous findings reported in the literature \cite{BavotaCLPOS12}.

A close-related work to ours is the one by Ferreira \etal~\cite{Ferreira:icse2018}, who conducted a study on five Java projects, 20,689 refactoring actions and 1,033 bug reports, looking at the distance between the commit in which the refactoring was performed and the commit in which the bug emerged in the refactored code element. They found that (i) many bugs are introduced in the refactored code as soon as the first immediate change is made on it, and (ii) code elements affected by refactoring actions performed in conjunction with other changes (\ie floss refactoring) are more prone to have bugs compared to root-canal refactoring actions. 

Indeed, we used a similar toolchain (\eg \tool{RMiner} to detect refactoring actions, SZZ to identify fix-inducing commits). However, the study design, the answered RQs, and the scale of the studies are different.

\section{Conclusions}
\label{sec:conclusions}
%%%%%%%%%%%%%%%%%%%%%%%%%%%%%%%%
%%%%%%%%%%%%%%%%%%%%%%%%%%%%%%%%

This paper reported a differentiated replication of a previous study by Bavota \etal \cite{BavotaCLPOS12}, using a different and up-to-date tool chain, finer granularity and more precise matching between refactoring actions and fix-inducing changes, and being conducted at a larger scale (103 projects in total).

The data extraction itself posed several challenges and highlighted important lessons for researchers conducting similar studies. First, carefully test the tool chain (including third-party tools) being used. Second, refrain to perform an indiscriminate, large-scale mining from GitHub. While previous studies already advised about the risks of mining GitHub \cite{KalliamvakouGBS16} and ranking projects by stars \cite{BorgesV18}, or provided means to identify a diverse and representative set of projects \cite{NagappanZB13}, for our study we found that only relying on a set of project belonging to a well-disciplined ecosystem (\ie the Apache Software Foundation projects) allowed us to have enough confidence to mine projects with a good linking between commits and issues and issue classification. 

The quantitative study results were surprising. Albeit the tool chain and the analysis methodology (\ie commit-level of granularity and matching of lines affected by refactoring actions and fix inducing changes) was completely different, and although we found how the size of a change played a significant role, results of the replicated study were generally confirmed, and the effect of refactoring appeared even more evident. Noteworthy, results hold both on the three systems analyzed in the original study and on a larger set including 100 additional Apache projects. These findings also support the observations reported in previous qualitative studies with developers \cite{Kim:fse2012}, indicating their concerns about possible bugs introduced in the refactoring process.

However, a deep, manual analysis on a sample of 384 fix-inducing changes overlapping with refactoring partially debunked the quantitative results, revealing that a quantitative analysis may ``scratch-the surface'' and miss details on how exactly the source code changed over time. At the same time, there is still a good proportion of cases in which refactoring actions indeed induce fixes, and there are recurring patterns in such cases. Often such recurring patterns highlight latent implications, \eg reverted changes might imply that some refactoring actions were not carefully planned. 

\new{The obtained results entail implications for both researchers and practitioners. As for researchers, the study highlights the need for better refactoring support, in particular for better planning/pondering it (\eg in the direction of identifying its possible impact \cite{ChaparroBMP14}), or automatically testing/verifying the change made, or further work in the direction of supporting refactoring review \cite{GeSWM17}. As for practitioners, this study warns them by pointing out that refactoring is only in theory behavior-preserving, therefore it must be planned with appropriate verification \& validation activities aimed at reducing its risks.}

We believe that our study, together with the previously published research on the same topic \cite{Weissgerber:msr2006,BavotaCLPOS12,Ferreira:icse2018}, provides substantial quantitative evidence of the relationship between refactoring and bug-fixing activities. However, we still see the need for more qualitative studies unveiling the mechanisms through which refactoring operations introduce bugs. Our future work will point in this direction.

The data and scripts used in our study are publicly available~\cite{replication}.
\section*{Acknowledgments}
\label{sec:ack}
This work has received funding from the European Research Council (ERC) under the European Union's Horizon 2020 research and innovation programme (grant agreement No. 851720).

\bibliographystyle{ACM-Reference-Format}
\bibliography{main}

\end{document}